\documentclass[prl,twocolumn,showpacs]{revtex4}
\usepackage{graphicx,graphics,color,times,slashed}
\begin{document}
\title{Realizing and Detecting the Haldane's Quantum Hall effect with Ultracold Atoms}

\author{L. B. Shao$^{1,2,3}$}
\author{Shi-Liang Zhu$^{1}$}\email{slzhu@scnu.edu.cn}
\author{ L. Sheng$^{2}$, D. Y. Xing$^{2}$}
\author{Z. D. Wang$^{3}$}
 \affiliation{$^1$Institute for Condensed Matter Physics and
Department of Physics, South China Normal
University, Guangzhou, China\\
$^2$National Laboratory of Solid
State Microstructure and Department of Physics, Nanjing
University, Nanjing, China \\
$^3$Department of Physics and Center
of Theoretical and Computational Physics, The University of Hong
Kong, Pokfulam Road, Hong Kong, China}



\begin{abstract}
We design an ingenious scheme to realize the Haldane's quantum Hall
model without Landau levels by using ultracold atoms trapped in an
optical lattice. Three standing-wave laser beams are used to
construct a wanted honeycomb lattice, where different on-site
energies in two sublattices required in the
model can be implemented through tuning the phase of one laser beam.
The staggered magnetic field is generated from the light-induced
Berry phase.
Moreover, we establish a relationship between the Hall
conductivity and the atomic density, enabling us to detect the
 Chern number with the typical
density-profile-measurement technique.
\end{abstract}

 \maketitle

The quantum Hall effect (QHE)~\cite{Klitzing} in two-dimensional
electron systems is one of the most peculiar quantum-mechanical
phenomena observed in nature. The QHE is usually associated with a
uniform external magnetic field, which splits the electron energy
spectrum into discrete Landau levels (LLs). When the Fermi energy
lies in the gap between two LLs, the Hall conductivity in units
$e^2/h$ is accurately quantized to an integer. The precise
quantization of the Hall conductivity was explained by
Laughlin~\cite{Laughlin} based upon a gauge invariance argument,
which is fundamental to the picture of edge states proposed by
Halperin~\cite{Halperin}. On the other hand, Thouless, Kohmoto,
Nightingale and Nijs (TKNN)~\cite{Thouless} interpreted the Hall
conductivity as the topological Chern number of the $U(1)$ bundle
over the magnetic Brillouin zone of the bulk states.

Twenty years ago, Haldane  showed in principle that a QHE may also
result from breaking of time-reversal symmetry without any net
magnetic flux through a unit cell of a periodic two-dimensional (2D)
system, where the electron states retain their usual Bloch state
character~\cite{Haldane}. In his work, Haldane constructed a
tight-binding model on a honeycomb lattice including a complex
second nearest-neighbor hopping integral. The honeycomb lattice
consists of two triangular sublattices $\bar{A}$ and $\bar{B}$ with
different on-site energies $M$ and $-M$, as shown in Fig.\ 1a. For
$M\neq 0$, the inversion symmetry is broken and the lattice
possesses the point group $C_{3\upsilon}$ symmetry. A periodic
vector potential $\mathbf{A}(\mathbf{r})$ is applied to the lattice,
given that the total magnetic flux through each unit cell vanishes,
i.e., the first-neighbor hopping integral $t$ is unaffected. The
second-neighbor hopping integral $t^{'}$ acquires a Peierls phase
factor $\mathrm{exp}(ie\int\mathbf{A}\cdot d\mathbf{r}/\hbar)$,
where the integration is along the hopping path. The Hamiltonian of
the model is written as
\begin{eqnarray}
H=&&\sum_{\langle l,j\rangle}(ta_{l}^{\dagger}b_{j}+H.c.)+\sum_{j}
M(a_{j}^{\dagger}a_{j}-b_{j}^{\dagger}b_{j})\nonumber\\
&&+\sum_{\langle\langle
l,j\rangle\rangle}t^{'}e^{i\varphi_{jl}}(a_{l}^{\dagger}a_{j}
+b_{l}^{\dagger}b_{j}),
\end{eqnarray}
where $a_{i}$ and $b_{i}$ are the annihilation operators on site
$R_{i}$ in sublattices $\bar{A}$ and $\bar{B}$, respectively.
$\varphi_{jl}$ is the accumulated Peierls phase from site $j$ to its
second neighbor $l$, which is assumed to take the form
$\varphi_{jl}=\pm\varphi$. The hopping directions for which
$\varphi_{jl}=+\varphi$ are shown in Fig.\ 1a. The most interesting
and unique feature of the model lies in
 that the phase of the system can be changed from a normal
insulator to a Chern insulator by the simulation of parity anomaly
~\cite{Haldane,deser,rjackiw,bardeen,semenoff,Novoselov,Zhang}.
However, it is extremely hard to realize the Haldane's model
experimentally in ordinary condensed matter systems because of the
unusual staggered magnetic flux assumed in the model.

 On the other hand, the technology of ultracold atoms in an optical lattice provides a
perspective approach to explore rich fundamental phenomena of
condensed matter physics~\cite{anglin,jaksch,duan}. In particular,
how to realize the QHE with cold atoms has attracted considerable
interest\cite{Cooper,Ho,Palmer,Sorensen,Viefers}. Nevertheless, the
atomic QHE has not been  observed yet, mainly due to challenges in
both the realization and detection of the atomic Hall effects.
Although an effective magnetic field for neutral atoms can be
simulated either by rotating the atoms\cite{Wilkin} or by
laser-induced Berry
phases\cite{Juzeliunas,Osterloh,zhusl1,Juzeliunas2006}, the strong
magnetic field region required for QHE  has not been reached yet in
experiments.  For the rotating method, the system is close to the
point at which the centrifugal potential cancels the external
harmonic trap, and the atoms may fly apart at the rotation speed
required by QHE~\cite{Ho,Viefers}. For the laser-induced Berry phase
approach, the cold atoms moving in a spatially varying laser field
feel an effective gauge
potential\cite{Juzeliunas,Osterloh,zhusl1,Juzeliunas2006}, but the
region of the strong uniform field is rather small for two typical
counterpropagating Gaussian laser beams. In addition, the detection
method for cold atoms is very different from that for condensed
matter systems; especially the widely used technique for QHE based
on the transport measurements is not workable for atomic QHE. In
this Letter, we design an ingenious scheme to realize the Haldane's
quantum Hall model without LLs by using ultracold atoms trapped in
an optical lattice. We work out a distinct method to construct the
honeycomb lattices that have different on-site energies by three
standing-wave laser beams.
Although it is still hard to achieve a strong homogenous magnetic
field required by the conventional QHE in atomic system, we may
evade this bottleneck since LLs are not necessary in this
unconventional QHE. We elaborate that the staggered magnetic field,
which is hard to generate in condensed matter systems, may be rather
easy to set up by other three standing-wave laser beams. In this
scenario, different on-site energies in two sublattices can be
easily adjusted through tuning the phase of one of the laser beams,
and thus the whole phase diagram\cite{Haldane} including the exotic
topological phase transition predicted by Haldane may be revealed
experimentally. Furthermore, based on $(2+1)$-dimensional
relativistic quantum mechanics calculations, we establish a direct
relationship between the Hall conductivity and the equilibrium
atomic density, such that the famous topological Chern number may be
experimentally detected with the standard density profile
measurement used in atomic systems\cite{anglin}.

\begin{figure}\label{fig1}
\includegraphics[totalheight=7.0cm,width=7.5cm]{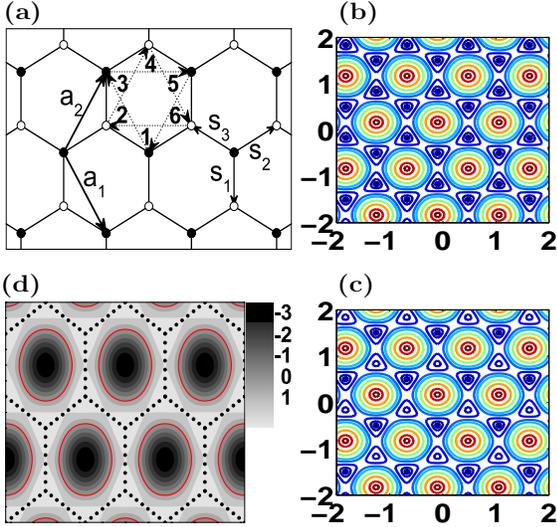}
\caption{(Color online) (a) Illustration of the Honeycomb lattice
structure of graphene, where open and solid circles represent
sites in sublattices $\bar{A}$ and $\bar{B}$.
$\mathbf{a}_{1}=(\frac{1}{2}a,-\frac{\sqrt{3}}{2}a)$,
$\mathbf{a}_{2}=(\frac{1}{2}a,\frac{\sqrt{3}}{2}a)$ are the unit
vectors of the underlying triangular sublattice. $s_{1}$, $s_{2}$
and $s_{3}$ are three vectors pointing from a $\bar{B}$ site to
its three nearest-neighbor sites.  (b) and (c) show the contours
of the potential V for $\chi=2\pi/3$ and $\chi=39\pi/60$,
respectively. The vertical (horizontal) axis represents
$yk_0^L/\pi$ ($xk_0^L/\pi$). (d) Contours of the magnetic field
defined by Eq.(\ref{Potential}). }
\end{figure}

Let us first consider single component fermionic atoms (e.g.,
$^{40}$K, $^6$Li, etc.) in a 2D honeycomb lattice\cite{duan,zhusl},
which can be realized by three detuned laser beams. A detuned
standing-wave laser beam will create a potential in the form
$V_{0}\sin^{2}(\mathbf{k}_{0}^{L}\cdot\mathbf{r})$, where $V_{0}$ is
the potential amplitude and $\mathbf{k}_{0}^{L}$ is the wave vector
of the laser.  To generate the honeycomb lattice with different
on-site energies in sublattices $\bar{A}$ and $\bar{B}$, the three
laser beams with the same wave length but different polarizations
are applied along three different directions: $\mathbf{e}_{y}$ and
$\frac{\sqrt{3}}{2}\mathbf{e}_{x}\pm\frac{1}{2}\mathbf{e}_{y}$,
respectively. The potential
is thus given by
$V= V_{0}[\sin^{2}(\alpha_+
+\frac{\pi}{2})+\sin^{2}(yk_{0}^L+\frac{\pi}{3})\nonumber+
\sin^{2}(\alpha_- -\frac{\chi}{2})],$
where $\alpha_{\pm}=\sqrt{3}xk_{0}^L/2 \pm yk_{0}^L/2$. The
potential contours are plotted in Fig. 1b and Fig. 1c. Under these
conditions, atoms are trapped at the minima of the potential,
forming a honeycomb lattice. An amazing feature here is that the
different site-energies of sublattices $\bar{A}$ and $\bar{B}$ is
controllable by the phase of laser beam $\chi$. For instance,  we
get exactly the honeycomb lattice with the same on-site energies
($M=0$) for  $\chi=\frac{2}{3}\pi$, as shown in Fig.1b, while  the
two sublattices have different on-site energies ($M\neq 0$) for
$\chi\neq\frac{2\pi}{3}$.



Now we elaborate how to simulate the staggered magnetic field in the
Haldane's model. Since the net flux per unit cell vanishes, the
vector potential applied to the lattice must be periodic. Such
magnetic fields can be created by Berry phase induced from two
opposite-travelling standing-wave laser beams~\cite{zhusl1}. For the
two laser beams with Rabbi frequencies
$\Omega_{1}=\Omega_{0}\sin(yk_{2}^L+\frac{\pi}{4})e^{ixk_{1}^L}$ and
$\Omega_{2}=\Omega_{0}\cos(yk_{2}^L+\frac{\pi}{4})e^{-ixk_{1}^L}$,
the effective gauge potential is generated as
$\mathbf{A}_1(\mathbf{r})=\hbar
k_{1}^L\sin(2yk_{2}^L)\mathbf{e}_{x}$~\cite{zhusl1}. Here,
$k_{1}^L=k^L\cos\theta$ and $k_{2}^L=k^L\sin\theta$ with $k^L$ the
wave vector of the laser and $\theta$ the angle between the wave
vector and the $\mathbf{e}_{x}$ axis. We emphasize that the choice
of wave vector $k_{2}^L$ of the laser beams must be a multiple of
$\frac{2\sqrt{3}\pi}{3a}$ in order to be commensurate with the
optical lattice. We take $k_{2}^L=\frac{2\sqrt{3}\pi}{3a}$. The
Peierls phases for the nearest-neighbor hopping in Fig.\ 1a are
$\varphi_{12}=\varphi_{61}=-\varphi_{34}=-\varphi_{45}=\varphi_{0}$
and $\varphi_{23}=\varphi_{56}=0$. For the next-nearest-neighbor
hopping integrals, which are integrated on a period of the vector
potential, the corresponding accumulated phases are
$\varphi_{13}=\varphi_{24}=\varphi_{46}=\varphi_{15}=0$, and
$\varphi_{35}=\varphi_{62}=\varphi$, where $\varphi= k_{1}^L
a\sin\frac{ak_{2}^L}{\sqrt{3}}$. Since the lattice has the symmetry
of point group $C_{3v}$, the vector potential $\mathbf{A}_{1}$ is
rotated by $\pm\frac{2}{3}\pi$ to obtain the other two vector
potentials. Then the total accumulated phases along the
nearest-neighbor directions are found to cancel out because of the
symmetry of honeycomb lattice. However, the total accumulated phases
for the next-nearest-neighbor hopping along the arrowed directions
of the dashed lines in Fig.\ 1a are just $\varphi$. Therefore, the
total vector potential and magnetic field can be written as
\begin{eqnarray}
\label{Potential}
\mathbf{A}&=&\hbar k_{1}^L[\sin(2yk_{2}^L)+\cos(\sqrt{3}xk_{2}^L)\sin(yk_{2}^L)]\mathbf{e}_{x}\nonumber\\
&\quad&-\sqrt{3}\hbar
k_{1}^L\sin(\sqrt{3}xk_{2}^L)\cos(yk_{2}^L)\mathbf{e}_{y}\\
\mathbf{B}&=&-2\hbar k_{1}^L k_{2}^L
[\cos(2yk_{2}^L)+2\cos(\sqrt{3}xk_{2}^L)\cos(yk_{2}^L)]\mathbf{e}_{z}.\nonumber
\end{eqnarray}
The contours of the magnetic field are plotted in Fig.\ 1d, in which
the red lines indicate where the magnetic field vanishes. The total
magnetic flux through each hexagon vanishes, as the Peierls phase
accumulated along its edges is zero. As a consequence, the total
Hamiltonian of this cold atomic system can be described by Eq.\ (1).
With Fourier transformation
$a_{j}=\frac{1}{\sqrt{N}}\sum_{\mathbf{k}}
e^{i\mathbf{k}\cdot\mathbf{R}_{j}}a_{k}$ and
$b_{j}=\frac{1}{\sqrt{N}}\sum_{\mathbf{k}}
e^{i\mathbf{k}\cdot\mathbf{R}_{j}}b_{k}$, the Hamiltonian of the
system can be written by using ``spinors'' $(a_{k},b_{k})^{t}$ as
\begin{equation}
\label{Spinor}
 H_k=h_{0}(\mathbf{k})+h_{1}(\mathbf{k})\sigma^{1}
+h_{2}(\mathbf{k})\sigma^{2}+h_{3}(\mathbf{k})\sigma^{3},
\end{equation}
where $h_{0}(\mathbf{k})=2t'\cos\varphi\sum_{i}\cos(\mathbf{k}\cdot
\mathbf{a}_{i})$, $h_{1}(\mathbf{k})=t\sum_{i}\cos(\mathbf{k}\cdot
\mathbf{s}_{i})$, $h_{2}(\mathbf{k})=t\sum_{i}\sin(\mathbf{k}\cdot
\mathbf{s}_{i})$, and
$h_{3}(\mathbf{k})=M+2t^{'}\sin\varphi\sum_{i}\sin(\mathbf{k}\cdot\mathbf{a}_{i})$.
$\mathbf{s}_{i}$  ($i=1$, $2$, $3$) are the three vectors pointing
from a $\bar{B}$ site to its three nearest neighbors. The vector
$h(\mathbf{k})=[h_{1}(\mathbf{k}),h_{2}(\mathbf{k}),h_{3}(\mathbf{k})]$
is an effective magnetic field of the ``spinors''. The energy
spectra are $E=h_{0}(\mathbf{k})\pm|h(\mathbf{k})|$ and the energy
gap is $|h_{3}|$, where $h_3=h_{3}(K_{+})$( or $ h_{3}(K_{-}))$ with
$K_{\pm}=\pm\frac{4\pi}{3a}(1,0)$. If $h_{3}=0$, the conduction band
touches the valence band at two Dirac points $K_{+}$ and $K_{-}$.
The famous TKNN index or Chern number  for this system is given by
$C=\frac{1}{8\pi}\int_{1BZ}d^{2}k\epsilon^{\mu\nu}\hat{h}\cdot
(\partial_{k_{\mu}}\hat{h}\times\partial_{k_{\nu}}\hat{h})$~\cite{lee}
with $\hat{h}(\mathbf{k})$ as the unit vector of $h(\mathbf{k})$. It
is demonstrated that the gauge invariance $C$ can only take integer
values and the quantum Hall conductivity is proportional to the
Chern number~\cite{Thouless}. For the Haldane's model, different
phases of the system can be characterized by different values of
$C$. The phase diagram is depicted in Fig.\ (2a)\cite{Haldane}, in
which the solid line is the critical boundary between the normal
insulator with $C=0$ and the Chern insulator with $C=\pm 1$.
 It is notable that the
phase $\varphi=\pi \tan\theta$ can be controllable by simply
choosing the laser angle $\theta$, while the energy difference $M$
between sublattice can be tuned by the phase of laser beams. With
such controllability, it is promising to realize the exotic
topological phase transition between different Chern numbers in Fig.
2a.


We now turn to establish a direct connection between the topological
Cherm number and the atomic density, noting that the latter can be
detected with density profile measurements typically used in atomic
systems. We first develop a Green's function method to calculate the
atomic density. The system is actually described by a Dirac-like
Hamiltonian which can be obtained by expanding Eq.(\ref{Spinor})
around two Dirac points $K_{\pm}=\pm\frac{4\pi}{3a}(1,0)$. By the
substitution of $\mathbf{k}\rightarrow K_{\pm}+\mathbf{p}$, we have
$H_{\pm}=\mp\upsilon_{F}p_{1}\sigma^{1}-\upsilon_{F}p_{2}\sigma_{2}+
m_{\pm}\sigma_{3}$ at $K_{\pm}$, respectively, where
$\upsilon_{F}=\frac{\sqrt{3}at}{2}$ is the Fermi velocity and
$m_{\pm}=M\pm 3\sqrt{3}t^{'}\sin\varphi$. Under unitary
transformation $\sigma^{2}H_{-}\sigma^{2}$, we can write the
Hamiltonian in a more symmetric form
\begin{equation}
\label{Dirac}
H_{\pm}=-v_{F}p_{1}\sigma^{1}-v_{F}p_{2}\sigma^{2}+m\sigma^{3}\ ,
\end{equation}
where the notation $m=\pm m_{\pm}$ is introduced for simplicity
(i.e., we temporally omit subscripts $\pm$ below).

\begin{figure}[tbp]
\centering
\includegraphics[width=7.0cm,height=5.5cm,angle=0]{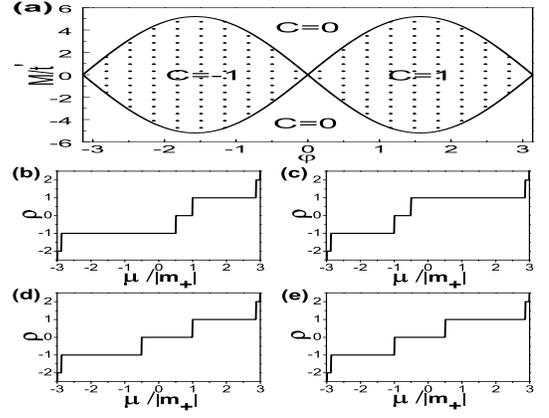}
\caption{ (a) Phase diagram of the system, where the phase
boundary (solid line) corresponds to $h_{3}=0$. The system behaves
like a normal insulator when $C=0$ and a Chern insulator when
$C=\pm 1$. (b)-(e) Charge density in units $\mathcal{B}/\phi_{0}$
for $|m_{-}|=0.5|m_{+}|$ and $e\mathcal{B}\hbar
v_{F}^{2}=4|m_{+}|$ as a function of normalized chemical potential
$\mu/\vert m_{+}\vert$ (corresponding to a rescaled atomic density
profile in a trap) for four different cases: (b) $m_{+}<0<m_{-}$,
(c) $m_{+}>0>m_{-}$, (d) $m_{+}<m_{-}<0$, and (e) $m_{+}>m_{-}>0$.
In (b) and (c) the shift of charge density at $\mu=0$ is present
so that the Hall conductivity is nonzero. In (d) and (e), the
shift of the atomic density is absent and so the Hall conductivity
vanishes.}\label{qubit}
\end{figure}

It is observed that the conductivity $\sigma_{xy}$ (Chern number) of
the system is related to the atomic density $\rho$ according to the
Streda formula $\sigma_{xy}=\partial\rho/\partial
\mathbf{\mathcal{B}}\vert_{\mu,T}$ once an additional uniform
magnetic field $\mathbf{\mathcal{B}}$ is applied. Such a magnetic
field can be simulated by rotating the optical lattice at a constant
frequency $\omega=e\mathbf{\mathcal{B}}/2m$. We choose the vector
potential as $\mathcal{A}_{0}=\mathcal{A}_{1}=0$ and
$\mathcal{A}_{2}=\mathcal{B}x$. By using the substitution
$\mathbf{p}\rightarrow\mathbf{p}+e\mathbf{\mathcal{A}}$ with $e>0$,
Eq.(\ref{Dirac}) can be solved in the real space. The eigenenergies
of the Hamiltonian can be obtained as~\cite{Haldane}
\begin{equation}\label{H}
E_{n}=\left\{
\begin{array}{ll}
 -m\mbox{sgn}(e\mathcal{B}) \text{\ \ \ \ \ \ }
 &n=0\\
 \pm\sqrt{m^{2}+2n\hbar v_{F}^{2}|e\mathcal{B}|}\text{ \ \ \ }
&n=1,2,3...
\end{array}
\right. ,
\end{equation}
and the degeneracy of each LL is ${|\mathcal{B}|}/{\phi_{0}}$ per
unit area with $\phi_{0}$
the flux quantum.
The density $\rho$ in terms of the Green's function for the Dirac
Hamiltonian is given by
\begin{equation}
\label{Green1} (\slashed{D}+m)G=1, \text{\ \ }\rho=-\mbox{Tr}[\gamma
^{0}G(\mathrm{x},\mathrm{x}^{^{\prime }})]|_{\mathrm{x}\rightarrow
\mathrm{x}^{^{\prime }}}
\end{equation}
where $\slashed{D}=\gamma^{\tau}D_{\tau}$, with
$D_{0}=\hbar\partial_{0}-\mu$, $D_{1}=\hbar v_{F}\partial_{1}$,
$D_{2}=\hbar v_{F}\partial_{2}+iev_{F}\mathcal{B}x$,
$\gamma^{0}=\sigma^{3}$, $\gamma^{1}=\sigma^{2}$, and
$\gamma^{2}=-\sigma^{1}$ in Euclidean space. From the standard Green
function approach, the atomic density is explicitly obtained as

\begin{eqnarray}
\rho_{\pm}&=&|\frac{\mathcal{B}}{\phi_{0}}|\mathrm{sgn}(\mu )
\left\{ \mathrm{int}\left[\frac{\mu ^{2}-m_{\pm}^{2}}{2\hbar v_{F}^{2}|e\mathcal{B}|}\right]+%
\frac{1}{2}\right\} \Theta (|\mu |-|m_{\pm}|)\nonumber \\
&&\pm\frac{\mathcal{B}}{2\phi_{0} }\frac{m_{\pm}}{|m_{\pm}|}\Theta
(|m_{\pm}|-|\mu |), \label{Density}
\end{eqnarray}
where $\Theta$ stands for the unit step function and
$\mbox{int}[x]$ means the largest integer less than $x$. The
second term of Eq. (\ref{Density}) is the atomic density induced
into vacuum as $\mu\rightarrow 0$ by the uniform magnetic field.
It is of parity anomaly since its corresponding Hall current is
independent of the magnetic field after multiplying the drift
velocity $\mathcal{E}/\mathcal{B}$, where $\mathcal{E}$ is the
electric field.

The total atomic density is given by the sum of the densities of the
two components $\rho=\rho_{+}+\rho_{-}$. At $\mu=0$, the Hall
conductivity at $\mathcal{B}=0$ can be obtained from the density by
using the Streda formula as $\sigma_{xy}=C\frac{e^2}{h}$, where
$C=\frac{1}{2}[\mbox{sgn}(m_{+})-\mbox{sgn}(m_{-})]$ is the Chern
number.
To show how to detect the Chern number of the system, we consider a
finite magnetic field $\mathcal{B}$. The calculated density $\rho$
in unit of $|e\mathcal{B}|/\phi_{0}$ is plotted as a function of the
normalized chemical potential $\mu/\vert m_{+}\vert$ (for
$|m_{-}|=0.5|m_{+}|$ and $e\hbar v_{F}^{2}\mathcal{B}=4|m_{+}|$) in
Fig.\ 2. It is essential that the spatial density profile $\rho (r)$
is uniquely determined by the function $\rho (\mu/\vert m_{+}\vert)$
in the local density approximation, which is typically well
satisfied for trapped fermions. Figures\ 2b-2e stand for four
different cases. The plateaus in the atomic density have one-to-one
correspondence to the plateaus in the Hall conductivity due to the
finite magnetic field $\mathcal{B}>0$. We here focus on $\mu=0$,
which is of our main interest. For $m_{+}<0<m_{-}$, which
corresponds to $C=-1$, the atomic density
$\rho=-\mathcal{B}/\phi_{0}<0$, as shown in Fig.\ 2b. For
$m_{+}>0>m_{-}$, which corresponds to $C=1$, the density
$\rho=\mathcal{B}/\phi_{0}>0$, as shown in Fig.\ 2c. For the other
two cases $m_{+}<m_{-}<0$ and $m_{+}>m_{-}>0$, $m_{+}$ and $m_{-}$
have the same sign, corresponding to $C=0$, and the density
$\rho=0$, as seen from Figs.\ 2d and 2e. Therefore, a simple direct
relation between the Chern number and the equilibrium atomic density
is established as
\begin{equation}\label{Chern_density}
 C=\rho\phi_{0}/\mathcal{B}.
 \end{equation}

The important relation (\ref{Chern_density}) actually provides us a
feasible way to experimentally detect the Chern number $C$ in
different phases. In the absence of $\mathcal{B}$, the density of
the cold atoms at $\mu=0$ is first measured, which is denoted as
$\rho_0$. Then the optical lattice is rotated to generate the
effective uniform magnetic field $\mathcal{B}$, and the new density
of the cold atoms $\rho_{1}$ is measured. If $\rho_{1}>\rho_{0}$,
the system is in a Chern insulator phase with Chern number $C=1$. If
$\rho_{1}<\rho_{0}$, the system is still a Chern insulator with
$C=-1$. However, if $\rho_{1}=\rho_0$, the system behaves like a
normal insulator with Chern number $C=0$. Since the density
difference is actually quantized in units $\mathcal{B}/\phi_{0}$,
the above method could be rather robust.

Finally, we briefly address an alternative approach to realize the
Haldane's QHE. The fermions we discussed are in the s-band of the
honeycomb lattice. As for fermions in the $p$-orbital bands, a
Haldane's quantum Hall model without LLs can also be implemented by
rotating each optical lattice site around its own center\cite{Wu}.
Nevertheless, how to detect such $p$-band QHE is still
desirably awaited.


In summary, we have shown that the Haldane's QHE model can be
realized by using ultracold atoms in an optical lattice. We have
established a relationship between the Hall conductivity and the
equilibrium atomic density, which provides a feasible way to
experimentally detect the Chern number $C$  in different phases.

This work was supported by the State Key Program for Basic
Researches of China (Nos. 2006CB921800, 2004CB619004, 2007CB925104,
2007CB925204, and 2009CB929504), the RGC of Hong Kong (Nos.
HKU7045/05P, HKU7049/07P and HKU7044/08P), the URC fund of HKU, NCET
and the NSFC under Grant Nos. 10429401, 10674049, and 10874066.

 {\it Note added} -- After this work was completed we
became aware that a relation between the density profile and the
Hall conductivity in conventional QHE was also addressed in
Ref.\cite{Zhai}. We thank Dr. H. Zhai for bringing our attention to
that work.


\end{document}